\documentclass[letter]{aa} % for the letters

\usepackage{natbib}
\bibpunct{(}{)}{;}{a}{}{,} % to follow the A&A style

\usepackage{graphicx}
\begin{document}
   \title{Magnetic Tension of Sunspot Fine Structures}

   %\subtitle{I. Overviewing the $\kappa$-mechanism}

   %\author{G. Wuchterl
%          \inst{1}
%          \and
%          C. Ptolemy\inst{2}\fnmsep\thanks{Just to show the usage
%          of the elements in the author field}
%          }
%
%   \institute{Institute for Astronomy (IfA), University of Vienna,
%              T\"urkenschanzstrasse 17, A-1180 Vienna\\
%              \email{wuchterl@amok.ast.univie.ac.at}
%         \and
%             University of Alexandria, Department of Geography, ...\\
%             \email{c.ptolemy@hipparch.uheaven.space}
%             \thanks{The university of heaven temporarily does not
%                     accept e-mails}
%             }

\author{P.~Venkatakrishnan and Sanjiv Kumar Tiwari}
\institute{Udaipur Solar Observatory, Physical Research Laboratory,
 Dewali, Bari Road, Udaipur-313 001, India\\
\email{pvk@prl.res.in}\\
\email{stiwari@prl.res.in}
}

%\date{Received September 15, 1996; accepted March 16, 1997}

% AA/2010/14786
%\abstract{}{}{}{}{}
% 5 {} token are mandatory

  \abstract
    %{}  %leave it empty if necessary
   {The equilibrium structure of sunspots depends critically on its
    magnetic topology and is dominated by magnetic forces. Tension
    force is one component of the Lorentz force which balances the gradient
    of magnetic pressure in force-free configurations.}
   {We employ the tension term of the Lorentz force to clarify the
    structure of sunspot features like penumbral filaments,
    umbral light bridges and outer penumbral fine structures.}
   {We compute vertical component of tension term of Lorentz force
   over two active regions namely NOAA AR 10933 and NOAA AR 10930 observed
   on 05 January 2007 and 12 December 2006 respectively. The former is a
   simple while latter is a complex active region with highly sheared
   polarity inversion line (PIL). The vector magnetograms used are obtained
   from Hinode(SOT/SP).}
   {We find an inhomogeneous distribution of tension with both positive and
    negative signs in various features of the sunspots.
    The existence of positive tension at locations of lower field strength and
    higher inclination is compatible with the uncombed model of the penumbral structure.
    Positive tension is also seen in umbral light bridges which could be indication
    of uncombed structure of the light bridge. Likewise, the upward directed tension
    associated with bipolar regions in the penumbra could be a direct confirmation
    of the sea serpent model of penumbral structures. Upward directed tension at
    the PIL of AR 10930 seems to be related to flux emergence.
    The magnitude of the tension force is greater than the force
    of gravity in some places, implying a nearly force-free configuration for these
    sunspot features.}
   {From our study, magnetic tension emerges as a useful diagnostic of the local equilibrium of the
   sunspot fine structures.}

   \keywords{Sun --
                magnetic fields --
                photosphere --
                sunspots
               }

   \maketitle
%_______________________________________________________________

\section{Introduction}

The equilibrium structure of sunspots is obviously dominated by
magnetic forces. Early researches on this problem dealt mainly with
the problem of the global equilibrium of sunspots \citep{meye77}.
The sunspot was modelled as a magnetic flux rope where the lateral
force balance was envisaged as a pressure balance between the
photospheric plasma pressure and the magnetic + plasma pressure
inside the flux rope \cite[e.g.,][]{chit63}. Since the magnetic
field is divergence free, the field lines of the flux rope must bend
back into the photosphere resulting in a ``closed'' field topology.
The resulting curvature of the field lines produces a magnetic
tension that should basically have a downward vertical component.
The model of \cite{meye77} also put constraints on the field line
curvature, since the equilibrium becomes unstable when the radius of
curvature is shorter than a certain value. With the availability of
high resolution magnetograms of sunspots, it has become clear that
the earlier models of sunspots might no longer be adequate to
explain the dynamical equilibrium of various fine structures seen in
the umbra as well as in the penumbra. One simple diagnostic of the
vertical equilibrium is the vertical component of the magnetic
tension, which can be determined from the lateral gradients of the
vector magnetic field. Information about the vertical component of
the magnetic tension was already found to be very useful, as e.g.
seen in the correlation of low tension force with large magnetic
shear in early vector magnetograms \citep{venk93} measured by the
MSFC vector magnetograph \citep{hagy82}. The magnetic tension
measurements at the polarity inversion lines underlying
filaments/prominences are like-wise very important since the
vanishing of magnetic tension at these highly sheared locations
makes the prominence structure extremely vulnerable to dynamical
instabilities, via thermal instabilities \citep{venk90}.

Modern observations of sunspots have revealed the existence of
different fine structures. In the umbra we have the umbral dots and
light bridges \cite{sobo89,sobo97,sobo97a}. In the penumbra we have
spines (stronger, more vertical field) wrapping around the
intraspines (weaker, more horizontal field) \citep{lites93,borr08}.
A few models have also been proposed to explain these structures
(e.g., uncombed model of \cite{sola93}, the gappy model of
\cite{spru06}).

In this paper, we investigate the height variation of the sunspot fine structure
using the calculations of the vertical component of the magnetic tension force.
The expression for computing the vertical component of tension force is
given in Section 2. In Section 3, we describe the data sets used. Section 4
describes the analysis and results. Finally in Section 5 we present our
conclusions.

\section{A Brief Description of the Magnetic Tension Force}

In any plasma with magnetic field {\bf B} and plasma pressure p,
the equation for magneto-hydrostatic equilibrium is given by \citep{parker79},
\begin{equation}\label{}
    (\nabla \times {\bf B}) \times {\bf B}/4\pi - \nabla p + \rho {\bf g} = 0
\end{equation}
where $\rho$ is the plasma density and {\bf g} is the acceleration
due to gravity. The first term in Equation 1 is the Lorentz force,
second term is the force due to plasma pressure and the last term is
the force on the plasma due to gravity. We can split up the Lorentz
force (say {\bf F}) in two terms as,
\begin{equation}\label{}
    {\bf F} = \frac{\bf (B \cdot \nabla) B}{4\pi} - \frac{\bf \nabla (B \cdot B)}{8\pi}
\end{equation}
The first term in this equation is the tension force (say {\bf T}). The second term represents
the force due to magnetic pressure. The vertical component of the tension term can be
expanded in terms of the horizontal derivatives of the magnetic field as:
\begin{equation}\label{}
{T_z} = \frac{1}{4\pi}[B_x \frac{\partial B_z}{\partial x} + {B_y} \frac{\partial B_z}{\partial y} -
B_z (\frac{\partial B_x}{\partial x} + \frac{\partial B_y}{\partial y})]
\end{equation}
where, the last component is drawn from the condition,
\begin{equation}\label{}
   {\bf \nabla \cdot B } = 0
\end{equation}
The utility of the tension force as a diagnostic of dynamical equilibrium has not
found much attention in the literature so far except in a restricted sense
\citep{venk90a,venk90,venk93}.
We have computed tension force using Equation 3 and expressed
it in the units of dynes/cm$^3$.

\section{Data Sets Used}

We have used the vector magnetograms of NOAA AR 10933 observed
on 05 January 2007 and NOAA AR 10930 observed on 12 December 2006.
These data sets are obtained from the Solar Optical
Telescope/Spectro-polarimeter
(SOT/SP: \citep{tsun08,suem08,ichi08,shim08}) onboard
Hinode \citep{kosu07}.

The Hinode (SOT/SP) data have been calibrated by the standard
``SP\_PREP'' routine developed by B. Lites and available in the
Solar-Soft package. The ``SP\_PREP'' determines the thermal shifts
in the spectral and slit dimensions and also applies the drift
corrections for calibrating the data from level0 to level1. The
prepared polarization spectra have then been inverted to obtain
vector magnetic field components using an Unno-Rachkowsky
\citep{unno56,rach67} inversion under the assumption of
Milne-Eddington (ME) atmosphere \citep{lando82,skum87}. We have used
the ``STOKESFIT" inversion code \footnote{The code has been
developed by T. R. Metcalf} which is available in the Solar-Soft
package. The latest version of the inversion code is used which
returns the intrinsic field strengths along with the filling factor.
The effect of polarimetric noise on the estimation of vector fields
is almost negligible \citep{tiw09a} and (Gosain, Tiwari and
Venkatakrishnan, 2010, ApJ, (sent)). The 180$^\circ$ azimuthal
ambiguity in our data sets have been removed by using acute angle
method \citep{harv69,saku85,cupe92}. The data sets used, have
spatial sampling of $\sim0.32$ arcsec/pixel observed in the ``fast
mode" scans of SOT/SP. A simple (NOAA AR 10933) and a complex (NOAA
AR 10930) sunspots are analyzed. The noise in the data has been
minimized in the similar way as done in \cite{tiw09e,tiw09b,venk09}.

\section{\textbf{Data Analysis and Results}}

\begin{figure}[h]
   \centering
   \includegraphics[width=0.48\textwidth]{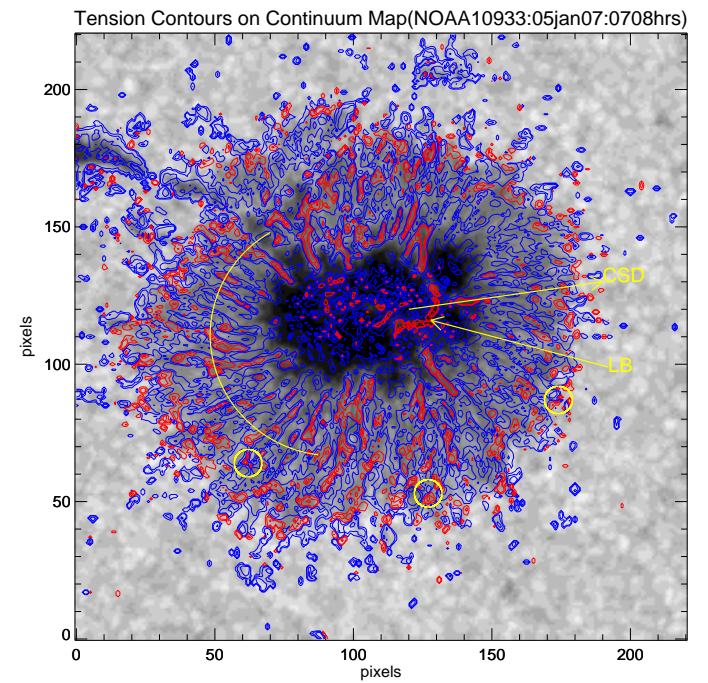}
      \caption{Contours of tension forces overlaid on the
continuum map of NOAA AR 10933 (S04E05). Blue (red) colors show negative (positive)
contours of $\pm1.2, \pm4, \pm12$ millidynes/cm$^3$.
The position of a light bridge(LB) and three examples of bipolar sea serpent regions are shown
by an arrow and circles respectively. An arc is shown for which the scatter plots of tension, field
strength and inclination are shown in Figure 2.
The heliocentric angle is $\theta$ = 8$^\circ$ and an arrow points towards the
center of the solar disk (CSD).}
         \label{F1}
\end{figure}

\begin{figure}[h]
   \centering
   \includegraphics[width=0.48\textwidth]{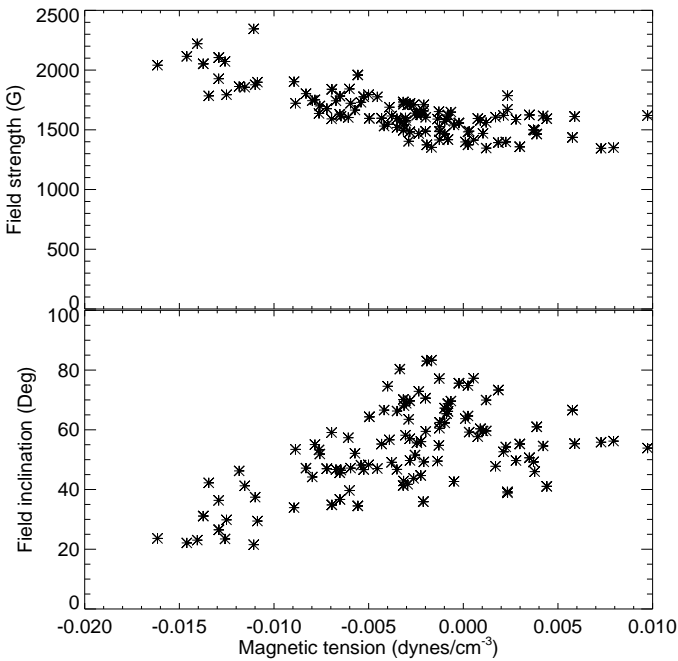}
      \caption{Upper panel: The scatter plot between magnetic tension and field strength
       along the arc shown in Figure 1. Lower panel: The scatter plot between
       magnetic tension and field inclination along the arc shown in Figure 1.}
         \label{F3}
\end{figure}

We calculate the vertical component of the tension force for all
locations in the sunspots. Figure 1 shows a continuum picture of the
NOAA AR 10933 overlaid by the contours of the magnetic tension
expressed in units of dynes/cm$^3$. The red contours depict positive
values, while the blue contours mark negative values. Several
interesting points can be noted. The most obvious result is the
existence of locations of positive tension, contrary to the
expectation from simple sunspot models having a ``closed" field
topology. Of further interest is the fact that these locations of
positive tension lie along linear structures in the penumbra, which
resemble the penumbral filaments. The magnitude of the tension is
very much comparable to the force due to gravity and exceeds the
solar gravity in some places. For a more detailed comparison, we
show in Figure A1 (online), a map of the magnetic field strength
overlaid with the contours of vertical tension. In Figure A2
(online), we present a map of the inclination overlaid with contours
of the magnetic tension. Here again, the locations of positive
tension largely coincide with regions of lower magnetic field
strength and larger magnetic inclination. In Figure 2, we show
scatter plots of field strength versus tension as well as
inclination versus tension. These scatter plots show a clear trend
of the positive tension being associated with weaker and more
inclined fields.

Apart from the penumbral features, we can also see several umbral
features including a light bridge at the western portion of the
umbra. Even in this case, the regions of positive tension coincide
with locations of weaker and more inclined fields. Finally, we
notice several locations of upward directed tension at the edges of
the filamentary penumbral structures. In Figure 3, we show the
contours of tension overlaid on the continuum map of NOAA AR 10930.
Apart from the similar structures of positive and negative tensions
as seen in the NOAA AR 10933 (Figure 1), the polarity inversion line
of NOAA AR 10930 also shows high magnitudes of
tension (10$^{-2}$ dynes/cm$^3$).  % ($\sim2g_{\odot}$ to $3g_{\odot}$).

\section{\textbf{Discussion and Conclusions}}

The variation of magnetic field parameters with optical depth can be
obtained \citep{coll94,west01,west01a,beck08} using inversion
schemes like SIR \citep{ruiz92,ruiz94,del96,del03}. With the
availability of high quality stokes profiles from the Solar Optical
Telescope \citep{tsun08,suem08,ichi08,shim08} aboard Hinode
\citep{kosu07}, these inversions are now possible with unprecedented
quality. In particular, this was used to demonstrate the variation
of magnetic field of penumbral filaments as a function of the
optical depth \citep{borr08}, which showed the wrapping of the spine
around the intraspine. An inspection of Figure 2 of \cite{borr08}
shows that the magnetic field lines can be inferred to have an
upward directed curvature above the intraspine. In the case of our
vector magnetograms, the inversions provide the magnetic parameters
at a single optical depth in the atmosphere. It has been recognized
that the simple inversions based on optical depth independent
magnetic parameters are consistent with a weighted mean of the
optical depth dependent inversions like SIR \citep{west98}. We can
thus conjecture that our locations of positive tension are
manifestations of the upward directed field curvatures that would be
seen in SIR inversions. This conjecture must be verified with
detailed SIR inversions over the entire sunspot, which will be
presented in a future detailed paper. For the present, we can only
note that the locations of positive tension in the penumbra appear
to be in the regions of lower field strength and larger inclination,
which are the properties of the intraspine. We can thus conclude
that the lateral variation of the magnetic field, as inferred from
simplistic inversions of the stokes profiles, appear to show an
upward directed tension, which may be in fact located above the
intraspine.

Likewise, the upward directed tension seen in light bridges could
well be due to a similar bending of field lines around a more
horizontal field \citep{sankar07}. Finally, the positive tension
seen at the edges in bipolar sections of the filamentary penumbral
structures could well be a signature of the ``sea-serpent"
\citep{sain08} where the penumbral fields dip into the photosphere
due to hydrodynamic forces caused by convection.

Apart from the sign, the magnitude of the vertical tension also has
some very interesting implications. As mentioned earlier, the
magnitude of tension attains values upto 10$^{-2}$ dynes/cm$^3$ in
some places. This value is comparable to the force of gravity which
ranges from 5$\times10^{-3}$ dynes/cm$^3$ for quiet Sun density to
10$^{-2}$ dynes/cm$^3$ for umbral densities. This must mean that
non-magnetic forces alone will not be able to balance this tension
force. The only other force that can match this force is the
gradient of magnetic pressure. As an example, we can use the values
of spine ($\sim 2000G$) and intraspine ($\sim 1000G$) of
\cite{borr08} and a scale height $\Lambda$ of $\sim$100 km for the
vertical variation of field strength, to estimate a downward
magnetic pressure gradient ($\frac{\Delta B^2}{8\pi\Lambda}$) of
$\sim 1.2\times10^{-2}$ dynes/cm$^3$. This value is of the same
order as that of magnetic tension obtained in our analysis. It is
well known that configurations where the magnetic tension balances
the magnetic pressure gradient are the so-called force-free
configurations. Thus, the use of high resolution magnetograms leads
to the important result that the existence of upward magnetic
tension in sunspot penumbral fine structure could be a manifestation
of uncombed magnetic field in {\it force-free equilibrium}.

Since the observations refer to values at a given optical depth, the
resultant magnetogram is not at a single geometrical height. What
could be the effect of this corrugation in the height levels? One
can estimate that the relative amplitude of tension fluctuations
($\Delta T/ T$) will be proportional to the relative amplitude in
the field variations ($\Delta B/ B$). Assuming a vertical gradient
of magnetic field of 1G/km in the penumbra and a corrugation
amplitude of 100 km for the optical depth, we obtain $\Delta B$ of
$\sim$ 100 G. Knowing that the penumbral fields are $\sim$ 1000 G,
we can estimate the relative field variation as 10\%, which then is
also the estimate for the relative variation in the tension force.
The observed spatial variation in tension is much greater than this
estimated artefact in tension fluctuations.

\begin{figure}[h]
   \centering
   \includegraphics[width=0.48\textwidth]{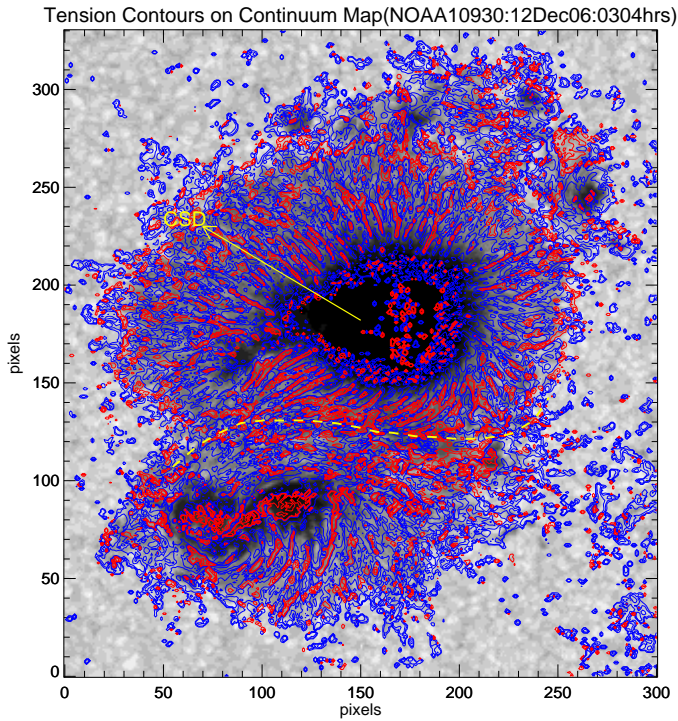}
      \caption{Contours of tension forces overlaid on the
continuum map of NOAA AR 10930 observed on 12 December 2006 at
an heliocentric angle of $\theta$ = 31$^\circ$.
Blue (red) colors show negative (positive)
contours of $\pm1.2, \pm4, \pm12$ millidynes/cm$^3$.
Polarity inversion line is shown by dashed yellow line. An arrow
points towards the center of the solar disk (CSD).}
         \label{F4}
    \end{figure}

%One interesting conclusion arises while considering the vertical balance
%of forces. If the sunspot magnetic field is indeed force-free,
%then the locations of upwardly directed tension will also be regions
%of downwardly directed magnetic pressure gradient.
%Then, magnetic pressure will increase with height.
%This is consistent with the location of positive tension in relatively weaker
%fields. This trend cannot be maintained at all heights, since magnetic
%field must vanish asymptotically at large heights.
%Thus, there will be a certain height at which the magnetic tension and magnetic
%pressure will separately vanish. This will imply a localized region of uniform
%magnetic field. Such a configuration will naturally arise above the intraspine
%when the field folds around the intraspine and becomes ``uncombed".
% This can be balanced by
%upward magnetic tension of a similar order. Our own results show similar values
%of positive magnetic tension at some places in sunspot structures.
%Thus, we also arrive at the conclusion that a combination of positive tension
%and force-free condition will result in an uncombed structure of the penumbral
%field which is consistent with the results obtained by SIR inversions of
%\cite{borr08}.

We must remember that the 180 degree ambiguity is not an issue
whenever the transverse magnetic vector makes an angle less than 90
degrees with the potential field. In all such cases, the ambiguity
is properly resolved by the acute angle method. The NOAA AR 10933 is
away from the polarity inversion line and is not highly sheared
\citep{tiw09b,venk09}. We note that the locations of positive
tension discussed thus far all have transverse vectors which are not
perpendicular to the potential fields. Hence they have no problem as
far as the acute angle method is concerned. We do see however that
this will be an issue for highly sheared portions of the magnetic
field. We normally expect low values of magnetic tension at such
highly sheared regions. But in the case of AR 10930, there is strong
magnetic tension seen in the highly sheared regions at the polarity
inversion line (Figure 3). In such cases, we need to have a careful
look at the 180 degree ambiguity resolution before we can decide on
the sign of magnetic tension.
%On the other hand, existence of positive tension at polarity inversion
%lines will require extremely innovative ways of arriving
%at the magnetic topology, e.g., highly twisted helical field lines crossing the
%polarity inversion line.
However, the upward tension could indeed be a signature of the
forces that drive flux emergence \citep{lites97}. This conjecture is
consistent with the copious emergence of flux observed on 12 Dec
2006 at the polarity inversion line (PIL) of the AR 10930
\citep{schr08}. In contrast, the PIL of AR 10930 on 14 December 2006
in Figure A3 (online) shows a reduced value of the tension
corresponding to a time of negligible flux emergence. Thus, the
study of magnetic tension in various types of sunspot fine
structures promises to yield new and exciting information on the
equilibrium and dynamics of these structures.

\begin{acknowledgements}
We thank Professors E. N. Parker and S. M. Chitre for reading the
manuscript, and Professors B. C. Low and R. Moore for their valuable
suggestions for improving the manuscript. We thank an anonymous
referee for his/her useful suggestions and comments to improve the
manuscript. Hinode is a Japanese mission developed and launched by
ISAS/JAXA, with NAOJ as domestic partner and NASA and STFC (UK) as
international partners. It is operated by these agencies in
co-operation with ESA and NSC (Norway).
\end{acknowledgements}

%\bibliographystyle{aa}
%\bibliography{stiwari}

\begin{thebibliography}{43}
\expandafter\ifx\csname
natexlab\endcsname\relax\def\natexlab#1{#1}\fi

\bibitem[{{Beck}(2008)}]{beck08}
{Beck}, C. 2008, \aap, 480, 825

\bibitem[{{Borrero} {et~al.}(2008){Borrero}, {Lites}, \& {Solanki}}]{borr08}
{Borrero}, J.~M., {Lites}, B.~W., \& {Solanki}, S.~K. 2008, \aap,
481, L13

\bibitem[{{Chitre}(1963)}]{chit63}
{Chitre}, S.~M. 1963, \mnras, 126, 431

\bibitem[{{Collados} {et~al.}(1994){Collados}, {Martinez Pillet}, {Ruiz Cobo},
  {del Toro Iniesta}, \& {Vazquez}}]{coll94}
{Collados}, M., {Martinez Pillet}, V., {Ruiz Cobo}, B., {del Toro
Iniesta},
  J.~C., \& {Vazquez}, M. 1994, \aap, 291, 622

\bibitem[{{Cuperman} {et~al.}(1992){Cuperman}, {Li}, \& {Semel}}]{cupe92}
{Cuperman}, S., {Li}, J., \& {Semel}, M. 1992, \aap, 265, 296

\bibitem[{{del Toro Iniesta}(2003)}]{del03}
{del Toro Iniesta}, J.~C. 2003, Astronomische Nachrichten, 324, 383

\bibitem[{{Del Toro Iniesta} \& {Ruiz Cobo}(1996)}]{del96}
{Del Toro Iniesta}, J.~C. \& {Ruiz Cobo}, B. 1996, \solphys, 164,
169

\bibitem[{{Hagyard} {et~al.}(1982){Hagyard}, {Cumings}, {West}, \&
  {Smith}}]{hagy82}
{Hagyard}, M.~J., {Cumings}, N.~P., {West}, E.~A., \& {Smith}, J.~E.
1982,
  \solphys, 80, 33

\bibitem[{{Harvey}(1969)}]{harv69}
{Harvey}, J.~W. 1969, PhD thesis, University of Colorado, Boulder

\bibitem[{{Ichimoto} {et~al.}(2008){Ichimoto}, {Lites}, {Elmore}, {Suematsu},
  {Tsuneta}, {Katsukawa}, {Shimizu}, {Shine}, {Tarbell}, {Title}, {Kiyohara},
  {Shinoda}, {Card}, {Lecinski}, {Streander}, {Nakagiri}, {Miyashita},
  {Noguchi}, {Hoffmann}, \& {Cruz}}]{ichi08}
{Ichimoto}, K., {Lites}, B., {Elmore}, D., {et~al.} 2008, \solphys,
249, 233

\bibitem[{{Kosugi} {et~al.}(2007){Kosugi}, {Matsuzaki}, {Sakao}, {Shimizu},
  {Sone}, {Tachikawa}, {Hashimoto}, {Minesugi}, {Ohnishi}, {Yamada}, {Tsuneta},
  {Hara}, {Ichimoto}, {Suematsu}, {Shimojo}, {Watanabe}, {Shimada}, {Davis},
  {Hill}, {Owens}, {Title}, {Culhane}, {Harra}, {Doschek}, \& {Golub}}]{kosu07}
{Kosugi}, T., {Matsuzaki}, K., {Sakao}, T., {et~al.} 2007, \solphys,
243, 3

\bibitem[{{Landolfi} \& {Landi Degl'Innocenti}(1982)}]{lando82}
{Landolfi}, M. \& {Landi Degl'Innocenti}, E. 1982, \solphys, 78, 355

\bibitem[{{Lites} {et~al.}(1993){Lites}, {Elmore}, {Seagraves}, \&
  {Skumanich}}]{lites93}
{Lites}, B.~W., {Elmore}, D.~F., {Seagraves}, P., \& {Skumanich},
A.~P. 1993,
  \apj, 418, 928

\bibitem[{{Lites} \& {Low}(1997)}]{lites97}
{Lites}, B.~W. \& {Low}, B.~C. 1997, \solphys, 174, 91

\bibitem[{{Meyer} {et~al.}(1977){Meyer}, {Schmidt}, \& {Weiss}}]{meye77}
{Meyer}, F., {Schmidt}, H.~U., \& {Weiss}, N.~O. 1977, \mnras, 179,
741

\bibitem[{{Parker}(1979)}]{parker79}
{Parker}, E.~N. 1979, {Cosmical magnetic fields: Their origin and
their
  activity} (Oxford, Clarendon Press; New York, Oxford University Press, 1979)

\bibitem[{{Rachkowsky}(1967)}]{rach67}
{Rachkowsky}, D.~N. 1967, Izv. Krymsk. Astrofiz. Obs., 37, 56

\bibitem[{{Ruiz Cobo} \& {del Toro Iniesta}(1992)}]{ruiz92}
{Ruiz Cobo}, B. \& {del Toro Iniesta}, J.~C. 1992, \apj, 398, 375

\bibitem[{{Ruiz Cobo} \& {del Toro Iniesta}(1994)}]{ruiz94}
{Ruiz Cobo}, B. \& {del Toro Iniesta}, J.~C. 1994, \aap, 283, 129

\bibitem[{{Sainz Dalda} \& {Bellot Rubio}(2008)}]{sain08}
{Sainz Dalda}, A. \& {Bellot Rubio}, L.~R. 2008, \aap, 481, L21

\bibitem[{{Sakurai} {et~al.}(1985){Sakurai}, {Makita}, \& {Shibasaki}}]{saku85}
{Sakurai}, T., {Makita}, M., \& {Shibasaki}, K. 1985, MPA Rep.,
No.~212, p.~312
  - 315

\bibitem[{{Sankarasubramanian} \& {Hagenaar}(2007)}]{sankar07}
{Sankarasubramanian}, K. \& {Hagenaar}, H. 2007, Bulletin of the
Astronomical
  Society of India, 35, 427

\bibitem[{{Schrijver} {et~al.}(2008){Schrijver}, {De Rosa}, {Metcalf},
  {Barnes}, {Lites}, {Tarbell}, {McTiernan}, {Valori}, {Wiegelmann},
  {Wheatland}, {Amari}, {Aulanier}, {D{\'e}moulin}, {Fuhrmann}, {Kusano},
  {R{\'e}gnier}, \& {Thalmann}}]{schr08}
{Schrijver}, C.~J., {De Rosa}, M.~L., {Metcalf}, T., {et~al.} 2008,
\apj, 675,
  1637

\bibitem[{{Shimizu} {et~al.}(2008){Shimizu}, {Nagata}, {Tsuneta}, {Tarbell},
  {Edwards}, {Shine}, {Hoffmann}, {Thomas}, {Sour}, {Rehse}, {Ito},
  {Kashiwagi}, {Tabata}, {Kodeki}, {Nagase}, {Matsuzaki}, {Kobayashi},
  {Ichimoto}, \& {Suematsu}}]{shim08}
{Shimizu}, T., {Nagata}, S., {Tsuneta}, S., {et~al.} 2008, \solphys,
249, 221

\bibitem[{{Skumanich} \& {Lites}(1987)}]{skum87}
{Skumanich}, A. \& {Lites}, B.~W. 1987, \apj, 322, 473

\bibitem[{{Sobotka}(1989)}]{sobo89}
{Sobotka}, M. 1989, \solphys, 124, 37

\bibitem[{{Sobotka} {et~al.}(1997{\natexlab{a}}){Sobotka}, {Brandt}, \&
  {Simon}}]{sobo97a}
{Sobotka}, M., {Brandt}, P.~N., \& {Simon}, G.~W.
1997{\natexlab{a}}, \aap,
  328, 682

\bibitem[{{Sobotka} {et~al.}(1997{\natexlab{b}}){Sobotka}, {Brandt}, \&
  {Simon}}]{sobo97}
{Sobotka}, M., {Brandt}, P.~N., \& {Simon}, G.~W.
1997{\natexlab{b}}, \aap,
  328, 689

\bibitem[{{Solanki} \& {Montavon}(1993)}]{sola93}
{Solanki}, S.~K. \& {Montavon}, C.~A.~P. 1993, \aap, 275, 283

\bibitem[{{Spruit} \& {Scharmer}(2006)}]{spru06}
{Spruit}, H.~C. \& {Scharmer}, G.~B. 2006, \aap, 447, 343

\bibitem[{{Suematsu} {et~al.}(2008){Suematsu}, {Tsuneta}, {Ichimoto},
  {Shimizu}, {Otsubo}, {Katsukawa}, {Nakagiri}, {Noguchi}, {Tamura}, {Kato},
  {Hara}, {Kubo}, {Mikami}, {Saito}, {Matsushita}, {Kawaguchi}, {Nakaoji},
  {Nagae}, {Shimada}, {Takeyama}, \& {Yamamuro}}]{suem08}
{Suematsu}, Y., {Tsuneta}, S., {Ichimoto}, K., {et~al.} 2008,
\solphys, 249,
  197

\bibitem[{{Tiwari}(2009)}]{tiw09e}
{Tiwari}, S.~K. 2009, PhD thesis, Udaipur Solar Observatory/Physical
Research
  Laboratory, Mohanlal Sukhadia University, Udaipur

\bibitem[{{Tiwari} {et~al.}(2009{\natexlab{a}}){Tiwari}, {Venkatakrishnan},
  {Gosain}, \& {Joshi}}]{tiw09a}
{Tiwari}, S.~K., {Venkatakrishnan}, P., {Gosain}, S., \& {Joshi}, J.
  2009{\natexlab{a}}, \apj, 700, 199

\bibitem[{{Tiwari} {et~al.}(2009{\natexlab{b}}){Tiwari}, {Venkatakrishnan}, \&
  {Sankarasubramanian}}]{tiw09b}
{Tiwari}, S.~K., {Venkatakrishnan}, P., \& {Sankarasubramanian}, K.
  2009{\natexlab{b}}, \apjl, 702, L133

\bibitem[{{Tsuneta} {et~al.}(2008){Tsuneta}, {Ichimoto}, {Katsukawa}, {Nagata},
  {Otsubo}, {Shimizu}, {Suematsu}, {Nakagiri}, {Noguchi}, {Tarbell}, {Title},
  {Shine}, {Rosenberg}, {Hoffmann}, {Jurcevich}, {Kushner}, {Levay}, {Lites},
  {Elmore}, {Matsushita}, {Kawaguchi}, {Saito}, {Mikami}, {Hill}, \&
  {Owens}}]{tsun08}
{Tsuneta}, S., {Ichimoto}, K., {Katsukawa}, Y., {et~al.} 2008,
\solphys, 249,
  167

\bibitem[{{Unno}(1956)}]{unno56}
{Unno}, W. 1956, \pasj, 8, 108

\bibitem[{{Venkatakrishnan}(1990{\natexlab{a}})}]{venk90a}
{Venkatakrishnan}, P. 1990{\natexlab{a}}, in IAU Symposium, Vol.
142, Basic
  Plasma Processes on the Sun, ed. {E.~R.~Priest \& V.~Krishan}, 323

\bibitem[{{Venkatakrishnan}(1990{\natexlab{b}})}]{venk90}
{Venkatakrishnan}, P. 1990{\natexlab{b}}, \solphys, 128, 371

\bibitem[{{Venkatakrishnan} {et~al.}(1993){Venkatakrishnan}, {Narayanan}, \&
  {Prasad}}]{venk93}
{Venkatakrishnan}, P., {Narayanan}, R.~S., \& {Prasad}, N.~D.~N.
1993,
  \solphys, 144, 315

\bibitem[{{Venkatakrishnan} \& {Tiwari}(2009)}]{venk09}
{Venkatakrishnan}, P. \& {Tiwari}, S.~K. 2009, \apjl, 706, L114

\bibitem[{{Westendorp Plaza} {et~al.}(2001{\natexlab{a}}){Westendorp Plaza},
  {del Toro Iniesta}, {Ruiz Cobo}, \& {Mart{\'{\i}}nez Pillet}}]{west01a}
{Westendorp Plaza}, C., {del Toro Iniesta}, J.~C., {Ruiz Cobo}, B.,
\&
  {Mart{\'{\i}}nez Pillet}, V. 2001{\natexlab{a}}, \apj, 547, 1148

\bibitem[{{Westendorp Plaza} {et~al.}(1998){Westendorp Plaza}, {del Toro
  Iniesta}, {Ruiz Cobo}, {Martinez Pillet}, {Lites}, \& {Skumanich}}]{west98}
{Westendorp Plaza}, C., {del Toro Iniesta}, J.~C., {Ruiz Cobo}, B.,
{et~al.}
  1998, \apj, 494, 453

\bibitem[{{Westendorp Plaza} {et~al.}(2001{\natexlab{b}}){Westendorp Plaza},
  {del Toro Iniesta}, {Ruiz Cobo}, {Mart{\'{\i}}nez Pillet}, {Lites}, \&
  {Skumanich}}]{west01}
{Westendorp Plaza}, C., {del Toro Iniesta}, J.~C., {Ruiz Cobo}, B.,
{et~al.}
  2001{\natexlab{b}}, \apj, 547, 1130

\end{thebibliography}

\clearpage\Online\appendix
\section{Additional figures}

\begin{figure}[h]
   \centering
   \includegraphics[width=0.5\textwidth]{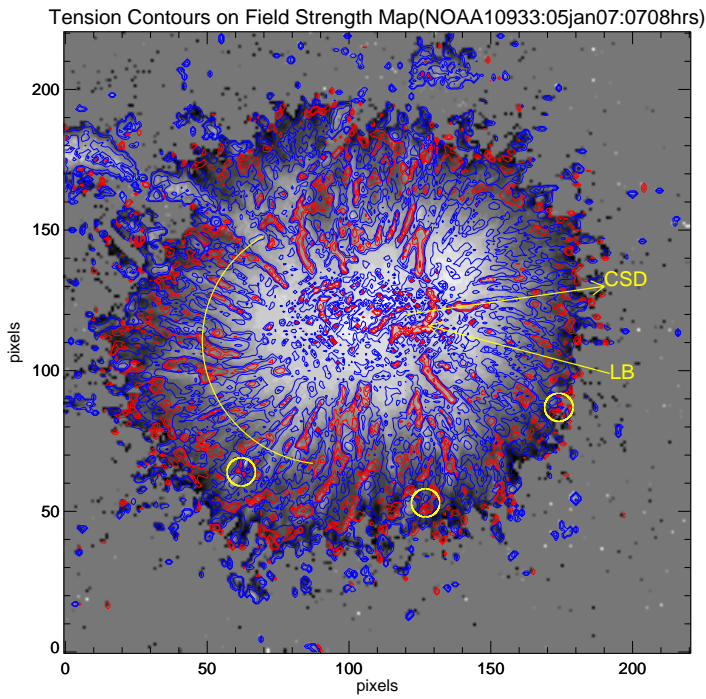}
      \caption{ Same as that of Figure 1, except that the continuum image is replaced by
      the field strength map of NOAA AR 10933.
      Contour levels are $\pm1.2, \pm4, \pm12$ millidynes/cm$^3$.}
         \label{F2}
   \end{figure}
\begin{figure}[h]
   \centering
   \includegraphics[width=0.5\textwidth]{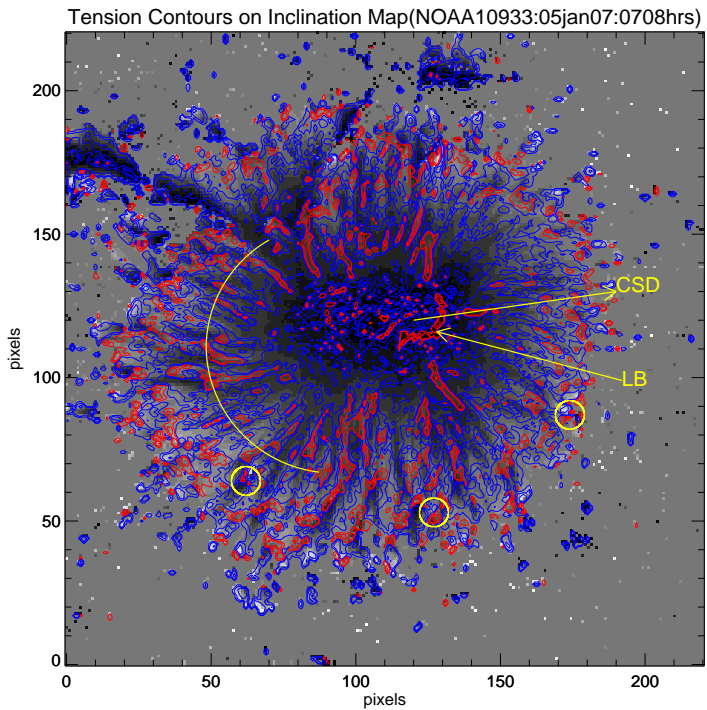}
      \caption{Same as that of Figure 1, except that the continuum image is replaced by
      the field inclination map of NOAA AR 10933.}
         \label{F3}
\end{figure}

\begin{figure}[h]
   \centering
   \includegraphics[width=0.5\textwidth]{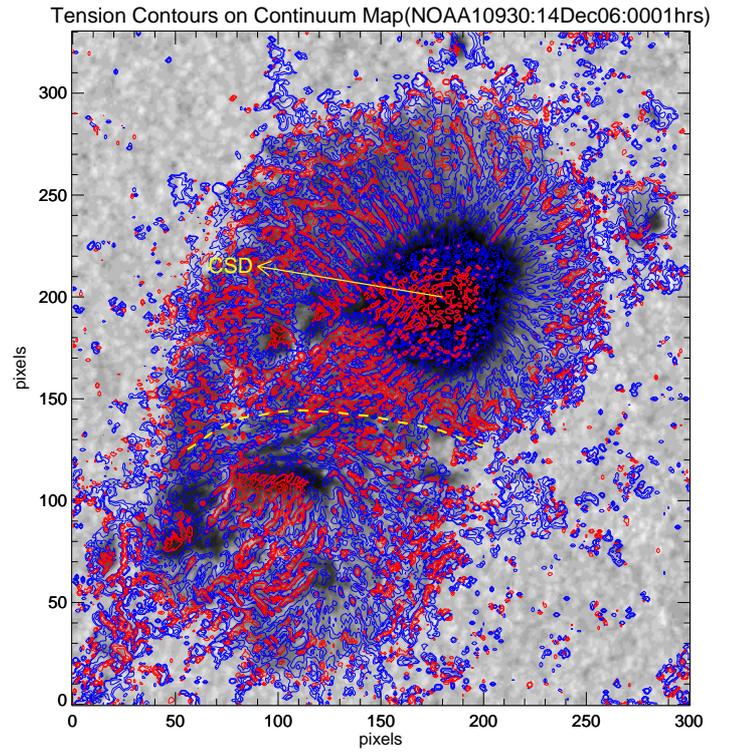}
      \caption{Same as that of Figure 5, except that this sunspot
      is observed on 14 December 2006 at an heliocentric angle of
      $\theta$ = 8$^\circ$.}
         \label{F4}
   \end{figure}

\end{document}